\documentclass[10pt, conference]{IEEEtran}

\usepackage[caption=false,font=footnotesize]{subfig}
\usepackage{amsmath,epsfig}
\usepackage{amssymb}
\usepackage{booktabs}
\usepackage{amsthm}
\usepackage{graphicx}
\usepackage{wrapfig}
\usepackage{rotating}
\usepackage{floatflt} 
\usepackage{paralist} 
\usepackage{algorithm} 
\usepackage{xcolor}
\usepackage{color}
\usepackage{epstopdf}
\usepackage[normalem]{ulem}
\usepackage{float}
\usepackage{flexisym}
\usepackage{cite}
\usepackage{colortbl}
\usepackage{multirow}
\usepackage{array}
\usepackage{arydshln} 
\usepackage{authblk}
\usepackage{siunitx}
\usepackage{mathtools}
\usepackage{tikz}
\usepackage{blindtext}
\usepackage{fancyhdr}

\newcolumntype{L}[1]{>{\raggedright\let\newline\\\arraybackslash\hspace{0pt}}m{#1}}
\newcolumntype{C}[1]{>{\centering\let\newline\\\arraybackslash\hspace{0pt}}m{#1}}
\newcolumntype{R}[1]{>{\raggedleft\let\newline\\\arraybackslash\hspace{0pt}}m{#1}}

\begin{document}

\title{\vspace{1cm}\huge {An Enclave-based TEE for SE-in-SoC in RISC-V Industry}}

\author{
  Xuanle Ren\\
  Computation Technology Lab \\
  DAMO Academy, Alibaba Group \\
  Shanghai, China \\
  Email: xuanle.rxl@alibaba-inc.com
  \and
  Xiaoxia Cui\\
  T-Head Semiconductor\\
  DAMO Academy, Alibaba Group\\
  Hangzhou, China\\
  Email: cxx194832@alibaba-inc.com 
}


\maketitle

\begin{tikzpicture}[remember picture,overlay]
\node[anchor=north east,yshift=-35pt,xshift=-15pt]%
    at (current page.north east)
    {\includegraphics[height=10mm]{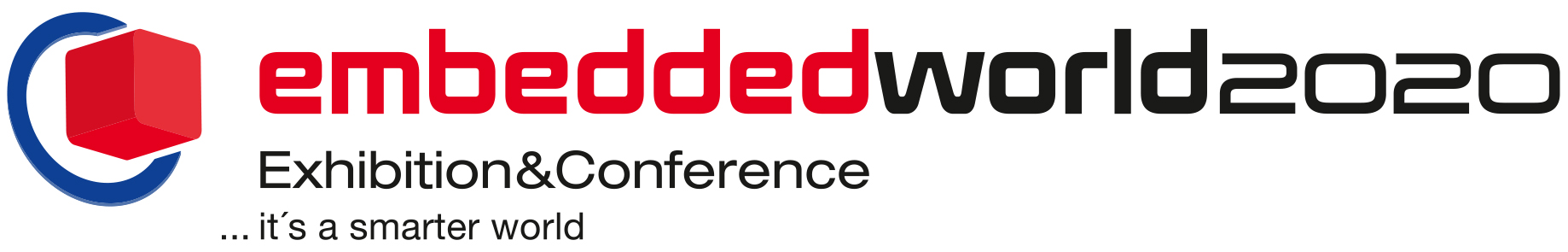}};
\end{tikzpicture}

\begin{abstract}

Secure Element (SE) in SoC sees an increasing adoption in industry. Many applications in IoT devices are bound to the SE because it provides strong cryptographic functions and physical protection. Though SE-in-SoC provides strong proven isolation for software programs, it also brings more design complexity and higher cost to PCB board building. More, SE-in-SoC may still have security concerns, such as malware installation and user impersonation. In this work, we employ TEE, a hardware-backed security technique, for protecting SE-in-SoC and RISC-V. In particular, we construct various enclaves for isolating applications and manipulating the SE, with the inherently-secure primitives provided by RISC-V. Using hardware and software co-design, the solution ensures trusted execution and secure communication among applications. The security of SE is further protected by enforcing the SE to be controlled by a trusted enclave and making the RISC-V core resilient to side-channel attacks.

\end{abstract}


\IEEEpeerreviewmaketitle
\section{Introduction} \label{sec:introduction}

During the last decades, we have seen a rapid growing of Internet-of-Things (IoT) devices. Thanks to the need of big data analysis, artificial intelligence (AI), and 5G. The IoT devices have been widely used in industry, automotive, smart cities, smart home, finance, and wearables. The number of connected IoT devices that are in use now exceeds 25 billion, and will be tripled by 2025 \cite{smart-home}. The IoT market is \$151B in 2018, and forecasted to grow to \$1,567B by 2025 \cite{IoT-analytics}. As another observation, IoT devices are not only used as an end device for collecting data, but also equipped with computation and even AI capability, meaning that more data process will be moved to IoT devices. This exacerbates design difficulty with the constraint of cost and causes security issues.

A Secure Element (SE) is a tamper-resistant platform capable of securely hosting applications and their confidential and cryptographic data in accordance with the rules and security requirements set by well-identified trusted authorities, like GlobalPlatform \cite{GP, OFL}. Secure Element (SE) is utilized in applications that require hardware cryptographic operations. The SE used to be a standalone chip, but recently it is more integrated within the SoC, avoiding communication between the SE and rest of the system exposed to attacks. Integrated SE has been widely implemented in many SoCs, such as mobile phones, SIMs, and smart cards.

With multiple applications being stored and their processes executed within a single device, it is essential to be able to host trusted applications and their associated credentials in a secure environment. In particular, authentication, identification, signatures, and PIN management are needed for different applications and require a protected environment to operate securely.

If not protected properly, attackers may compromise the cryptographic keys or dump sensitive data within the IoT devices \cite{IoT-attack}. More commonly, compromised IoT devices can be used as an entry point to access more devices in the network or even the cloud \cite{IoT-attack}. The security issue exacerbates considering that most IoT devices are not protected at all, while some of them are equipped with quite weak protection layer. A typical attack vector is to find vulnerabilities of the application code, using which the attacker can intrude the firmware layer. The attacker can then control the system by modifying the firmware, such as unlock the vehicles remotely.

A secure environment has been introduced by Trusted Execution Environment (TEE). A TEE is a secure area of a processor, which guarantees code and data loaded inside to be protected with respect to confidentiality and integrity. A typical TEE involves the Arm TrustZone \cite{TrustZone}. TrustZone divides on-chip resources into a secure world and an insecure world. The SE APIs are used by trusted applications (TAs) which are normally located in the secure world. All TAs can call the SE APIs, meaning that the communicated data between a TA and the SE can be observed or even modified by other TAs. This might threaten confidentiality and integrity of data. Other TEE solutions, including Intel SGX \cite{costan2016intel}, Keystone \cite{lee2019keystone}, MultiZone \cite{MultiZone}, and SiFive Shield \cite{Shield}, overcome this problem by partitioning the processor into more zones (or called enclaves) than two. In particular, Intel SGX isolates specific application code and data in memory, allowing user-level code to allocate enclaves. The enclaves are protected from processes running at higher privilege levels. This ensures security of the enclaves even if the OS is compromised. Keystone describes a TEE framework for RISC-V CPUs. Keystone relies on hardware primitives for memory isolation (i.e., physical memory protection, or PMP) and builds a runtime layer in each enclave, which acts as a software-programmable layer. The runtime, running in supervisor mode, provides functionality plugins for system call interfaces, libc support, in-enclave virtual memory management, self-paging, and more inside the enclave \cite{lee2019keystone}. The use of customizable runtime provides more design flexibility. Both SGX and Keystone are designed for server-level CPUs, thus more suitable for the cloud. MultiZone and SiFive Shield target RISC-V based IoT devices, both of which also consider SoC security rather than only the CPU. In particular, MultiZone uses the PMP to partition SoC resource for the CPU, and uses IOPMP to limit the access privilege of the other bus masters. In addition to PMP, the SiFive Shield marks the applications within different worlds. The marks determine the privilege of accessing the core, cache, interconnect, peripheral, and memory.

To this end, we propose XINE (eXecution-in-Enclave), a lightweight TEE framework for RISC-V based MCUs. The TEE, built on RISC-V, relies on its standard specification for PMP. The SE hardware is manipulated by a specially designed enclave, such that the cryptographic operations are isolated from applications. To call cryptographic functions, application enclaves need to switch the core to that specific enclave. In addition, we design another special enclave that stores runtime callable by other enclaves. This saves on-chip resource. All of these enclaves run in U-mode of the RISC-V, and they are switched by an enclave privilege arbitrator (EPA) running in M-mode. In addition, we implement a DMA controller for inter-enclave data transfer. Since the DMA allows effective data transfer without intervention of CPU, the DMA permission needs to be carefully designed to not violate security specifications. A new CSR is added to the CPU to arbitrate DMA transactions.

\section{XINE Overview} \label{sec:overview}

XiE is designed for an RISC-V based MCU that contains only M-mode (which can directly manage physical resources) and U-mode (where OS and user-space applications run).

\subsection{Design Principles}

The design should leverage the hardware primitives provided by RISC-V specifications. The RISC-V core defines multiple privilege levels, namely M-, S-, and U-mode. The CPU designer can decide how many privilege levels to implement according to the complexity of the system. The M-mode is usually regarded as trusted because it has the highest privilege level and can be fully verified. The higher privilege modes need to schedule applications of lower privilege modes, and also handle interrupts and exceptions. Physical memory protection (PMP) is another security primitive that provides hardware-based capability for memory isolation. It aims to partition memory address (including RAM, Flash, and GPIO) into multiple regions such that applications running in different regions cannot access each other.

The second principle is minimality. Many IoT devices are quite sensitive to cost, and therefore we need to make a balance between security and design cost. Two aspects are considered. First, the EPA is only used for secure boot, switching enclaves, and handling parts of exceptions/interrupts from enclaves. Second, the runtime is kept for only one copy instead of multiple copies. 

The third principle is scalability. The platform designer should be able to determine the number of enclaves. Note that an enclave refers to an isolated memory region, and the number of enclaves is limited by the number of the PMP setting. The RISC-V specifications suggest 16 entries of configuration \cite{RVSpecsII}, meaning that the memory address can be partitioned into up to 16 regions. 

Finally, the design should be extendable, meaning that more security primitives could be implemented without altering the current design fundamentally. For example, the design can be extended to 1) a RISC-V core with M-, S-, and U-mode, 2) a multi-core microprocessor, and 3) a more complex SoC, such as with bus masters other than the RISC-V core, and even with coprocessors. Some other security primitives include secure boot, security features for the CPU (anti-side-channel, anti-fault-injection), and memory encryption.

\subsection{Threat Model}

In this work, we assume that M-mode is trusted. This assumption is reasonable because the M-mode code, due to its small size, is fully verified at design time, and also it is measured with respect to its integrity during system boot (will be elaborated in Section~\ref{sec:xine-arch}). 

We define three assets that XINE aims to protect. The first asset is firmware that contains important information of the device and the system, such as the device ID and configuration of the system. Vulnerabilities of the IoT system are also fixed through updating the firmware. Thus, it is essential to avoid the firmware to be dumped or modified. The second asset is cryptographic secrets. Many IoT devices contain cryptographic modules. The generation, storage, and usage of keys need to be carefully protected. This also explains why integrated SE is replacing embedded SE. The third asset is sensitive data, such as personal information and user password.

Similar to the \cite{lee2019keystone}, we consider four types of attacks, namely, physical attack, software attack, side-channel attack and denial-of-service (DoS) attack. A physical attacker can intercept, modify, and/or replay signals that leave the chip package, but cannot observe, modify, or reverse engineer the components inside the chip package. A software attacker can control applications, the OS, and/or network communication. A side-channel attacker can observe the signals which are not aimed to be exposed to end-user but may cause information leakage. Typical side channels include cache timing, power analysis, and electromagnetic emission. DoS attacks refer to the malicious enclaves or the OS that refuse responding to the request from other enclaves, OS, or the EPA.

\begin{figure}[t!]
    \centering
  \subfloat[]{\includegraphics[width=.9\linewidth]{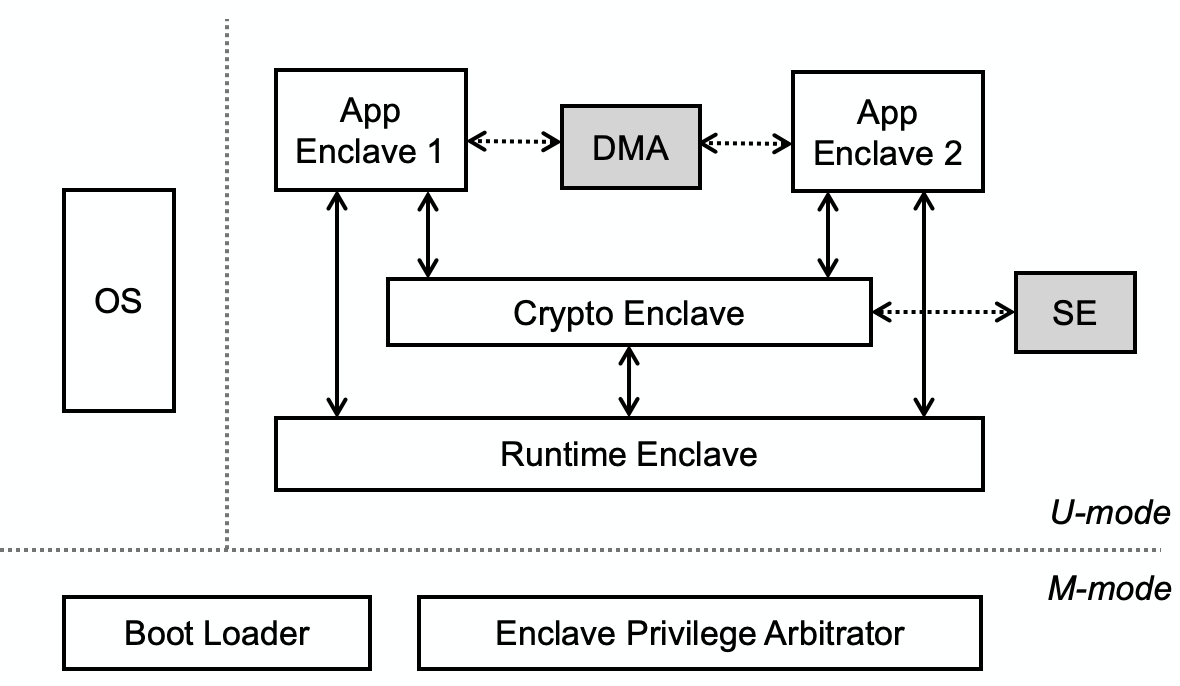}}
    \label{fig:overview}
  \subfloat[]{\includegraphics[width=.9\linewidth]{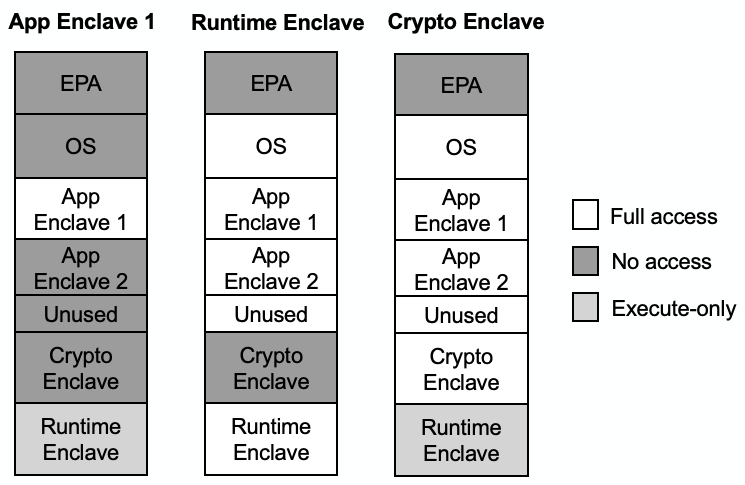}}
    \label{fig:privilege}
\caption{(a) Implementation of the TEE and (b) privileges of the enclaves.}
\label{fig:f1}
\end{figure}

\section{XINE Framework} \label{sec:xine-framework}

\subsection{XINE Architecture} \label{sec:xine-arch}

The overall design of XINE is shown in Fig.~\ref{fig:f1}(a). The TEE is implemented in a RISC-V core with U- and M-mode. The Enclave Privilege Arbitrator (EPA) runs at the M-mode, while the OS and the enclaves run at the U-mode. As shown in Fig.~\ref{fig:f1}(a), the TEE contains a variety of enclaves, namely a crypto enclave, a runtime enclave, and multiple app enclaves.

\subsubsection{Enclave Privilege Arbitrator} \label{sec:epa}

The M-mode code is used for accessing hardware, configuring control and status registers (CSRs), and initiating the bootloader. The EPA, running in M-mode, can switch execution between the OS and enclaves, and can also configure the PMP entries for an enclave, which guarantees isolation \cite{RVSpecsII}.

The PMP unit provides per-hart\footnote{The RISC-V defines the hart as a hardware thread.} M-mode control registers to allow physical memory access privileges (read, write, and execute) to be specified for each physical memory region. The memory region specified by each PMP entry consists of two sets of CSRs, namely an address register that specifies the address of the corresponding memory region, and a configuration register that specifies the access privilege and the mode of address matching. PMP checks are applied to all accesses when the hart is running in the S- or U-mode. Different from TrustZone where bus slaves (e.g., on-chip RAM/Flash, off-chip DDR, and peripherals) are gated by a configuration module, a PMP violation will be trapped at the processor.

\begin{figure}[t!]
\centering
\includegraphics[width=1\linewidth]{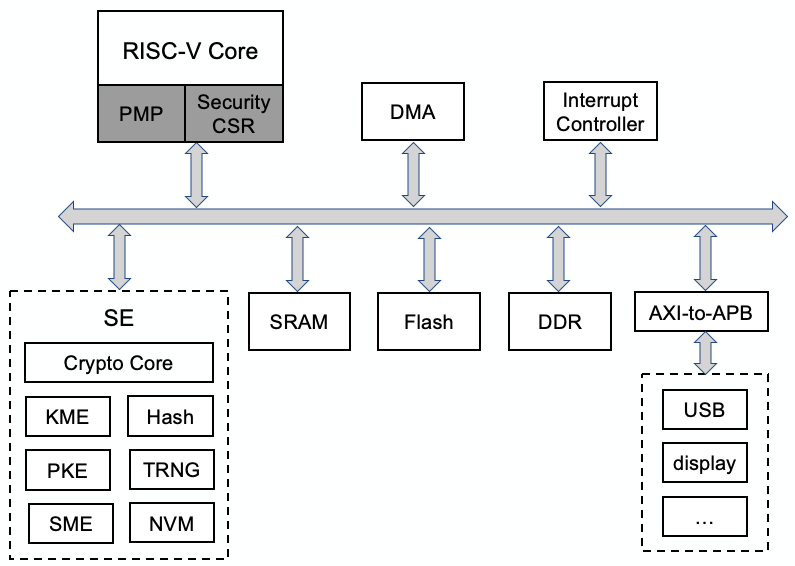}
\caption{The SoC contains two RISC-V cores, one for executing applications and the other for operating the SE.}
\label{fig:soc}
\end{figure}

\subsubsection{Enclaves} \label{sec:enclaves}

PMP-enforced isolation not only protects sensitive applications, but also jails malicious applications. The enclaves that encapsulate applications are named App Enclaves (AEs). In addition, we define two special enclaves, namely, a Crypto Enclave (CE) and a Runtime Enclave (RE).

The CE has exclusive control over the SE. Fig.~\ref{fig:soc} shows a typical SE which composed of a RISC-V core, cryptographic accelerators, a true random number generator (TRNG) and a piece of eFuse for storing keys. Different from traditional designs, the SE contains a separate RISC-V core. The RISC-V core, controlled exclusively by the CE, communicates with the main system via a mailbox. For instance, to encrypt data, the CE places the plain data in the mailbox (located in the CE), and then the RISC-V core reads the data and encrypts them using the SE hardware. Once completed, the RISC-V core places the encrypted data in the mailbox and notifies the CE to pick. The mailbox is only accessible to the CE, but not to the APs, which prevents the crypto message of AE-1 to be dumped or modified by AE-2.

When an AE needs to use cryptographic services provided by the CE, the AE needs to request the EPA to transfer control to the CE. Since the CE has full privilege to the AE region, the CE can run the requested cryptographic operation on the message of the AE directly. When the operation is completed, the CE writes the results in the address specified by the AE. A set of SE APIs is to be designed for supporting these operations.

The RE stores the runtime (such as the standard libc) that is callable by all AEs. As revealed in Fig.~\ref{fig:f1}(b), AEs can execute (but not read or write) the RE, which guarantees the confidentiality and integrity of the RE. Note that calling a RE function does not need to transfer control to RE. The use of RE avoids the runtime to be stored in each AE, thus resulting in less usage of on-chip resource.

\subsubsection{Inter-Enclave Communication} \label{sec:dma}

The PMP partitions memory region into different parts that are mutually isolated, but sometimes the enclaves may need to transfer data between each other. This can be achieved through a mailbox. In particular, the OS assigns a memory region that is accessible to both the source and the destination enclaves. The source enclave puts the message into the mailbox if the mailbox is not full, while the destination enclave gets the message from the mailbox if the mailbox is not empty. The size of the mailbox and the message format are pre-defined. This mechanism is efficient for data transfer between cores that are running in parallel because the source core and the destination core can frequently check the availability of the mailbox. In this work, the MCU has only one logic core. Frequently checking the mailbox may result in too many switches between the source and the destination enclaves.

\begin{figure}[t!]
\centering
\includegraphics[width=1\linewidth]{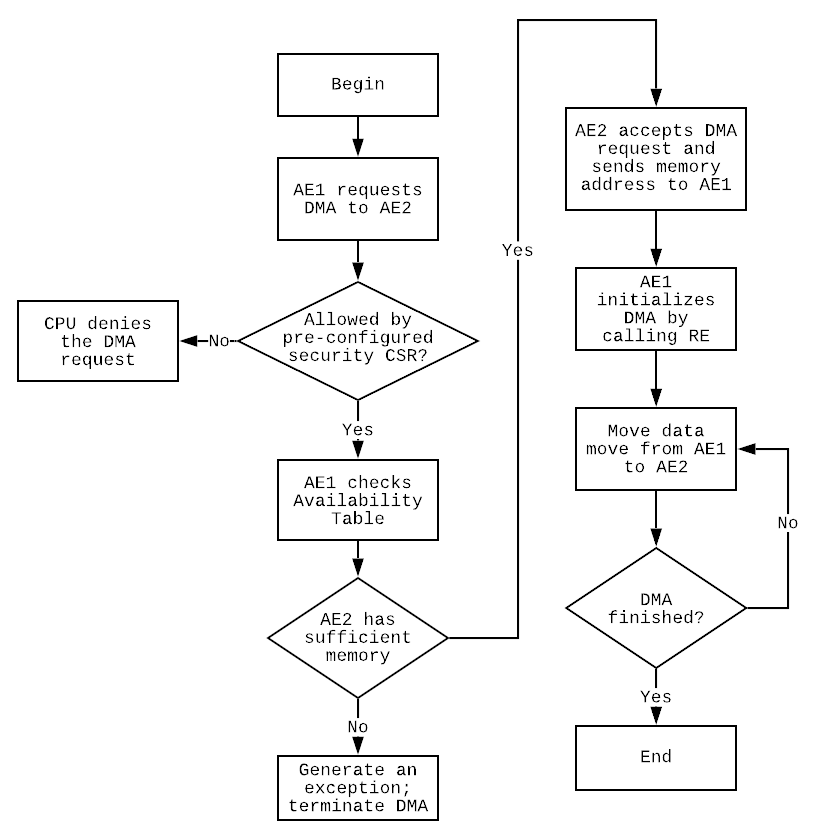}
\caption{Access control flow of the DMA.}
\label{fig:dma}
\end{figure}

To accelerate data transfer between AEs, we implement a direct memory access (DMA), as shown in Fig.~\ref{fig:soc}. DMA allows data to be transferred without intervention of the CPU, thus significantly improving data throughput. However, since the DMA unit is not a part of the CPU, it cannot tell if a DMA request is legitimate or not. To address this problem, we add a special security CSR within the CPU to check the legitimacy of a DMA request. Let us consider an example that AE-1 requests to move data to AE-2 (Fig.~\ref{fig:dma}). First, AE-1 makes a DMA request indicating the destination enclave and the data size. The request is checked by the security CSR. If the DMA transfer is allowed by the CSR, then AE-1 checks a special table (named Availability Table) which records the available memory space of AE-2. Only if AE-2 has enough memory space to receive the data, then the DMA controller starts the transfer. Here, the Availability Table, accessible to all AEs, is updated every time when an AE exits its execution. In addition, to avoid arbitrary DMA requests (by malicious applications), AEs can only request to move its data to other enclaves, but not able to request data from other enclaves.

\begin{figure}[t!]
\centering
\includegraphics[width=1\linewidth]{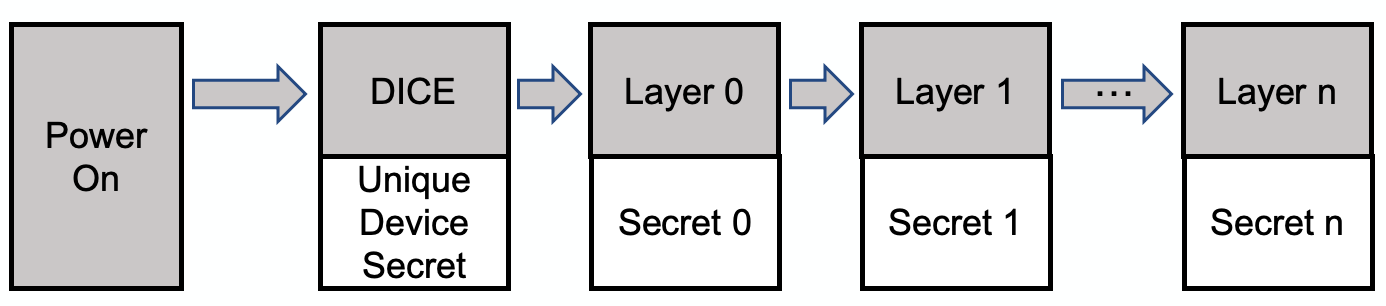}
\caption{Secure boot using DICE \cite{DICE-Flash}.}
\label{fig:dice}
\end{figure}

\subsubsection{Secure Boot} \label{sec:boot}

Secure boot requires not only a hardware root-of-trust that is not accessible to any user, but also a trusted boot chain where each step of boot code is measured for integrity. We employ Device Identifier Composition Engine (DICE), a lightweight architecture more suitable for IoT devices. The DICE provides hardware-based identity and attestation, as well as sealing, data integrity, device recovery and update \cite{DICE-GP}.

Fig.~\ref{fig:dice} shows the boot process using DICE. The DICE has exclusive access to the unique device secret (UDS). Once powered on, the DICE combines the UDS and the first mutable code (and optionally hardware state and configuration data), and computes the corresponding measurement. This layer results in a Compound Device Identifier, which supplies as the input to the next layer. This process continues until the whole boot chain is completed. In our work, XINE boots from the EPA, and then the CE and the RE. Each component in the boot chain needs to be measured and signed. If the measurement does not match the expected hash value, then the system fails to boot.




\subsection{Enclave Lifecycle} \label{sec:lifecycle}

\begin{figure}[th]
\centering
\includegraphics[width=.6\linewidth]{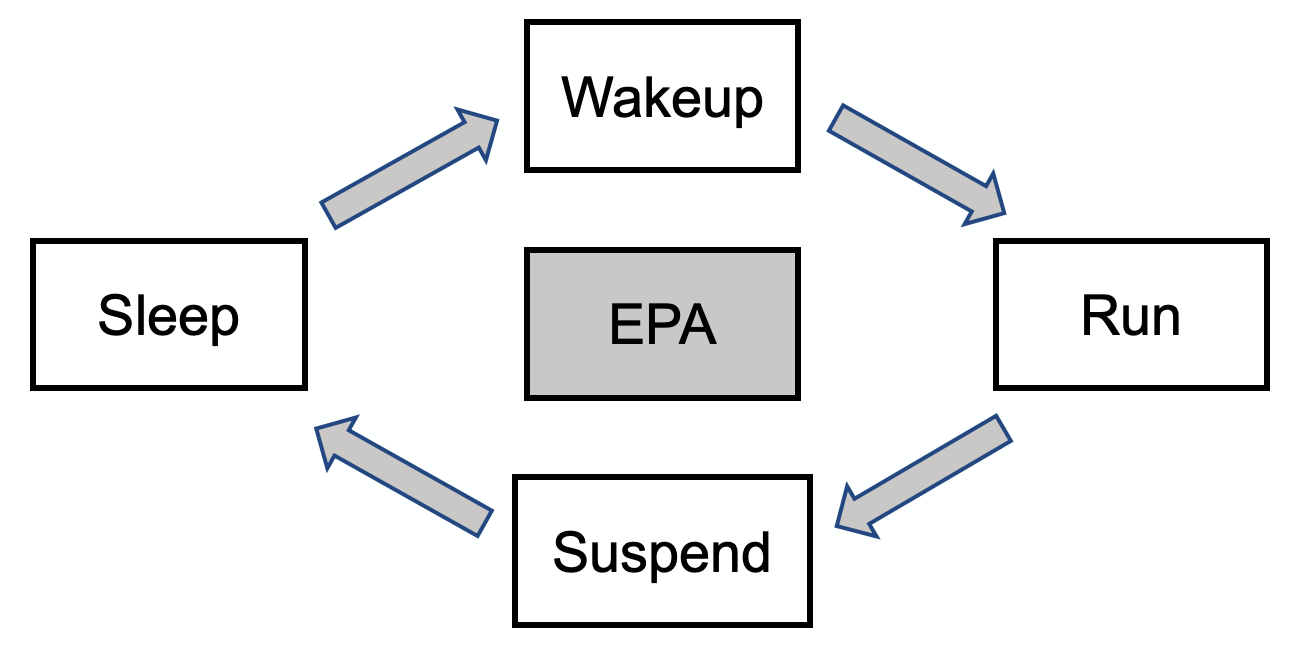}
\caption{The lifecycle of an App Enclave.}
\label{fig:lifecycle}
\end{figure}

Fig.~\ref{fig:lifecycle} shows the lifecycle of an AE. In the MCU, each AE is assigned with a static memory region enforced by the PMP. The code and data are measured before provisioned to the corresponding AEs. An AE sleeps if the PMP is not switched to its configuration entry and therefore the core is not executing its code. When the control is transferred to the AE, the corresponding PMP entry is set up, and the context of the AE is recovered. We name this process as a wakeup. Then the code within the AE starts to run from the specified entry point. The AE becomes suspended if one of the following events happens, namely 1) the AE requests to transfer the control to another AE or to the CE, and 2) an interrupt or an exception happens. In these situations, the CPU saves the context, and enters M-mode to switch the control to the intended enclave. 

\subsection{Security Analysis} \label{sec:analysis}

During enclave execution, the enclave only has access to its own memory region, and any illegal access will be refused by the PMP. In addition, the use of the security CSRs controls the privilege of DMA, such that illegal DMA requests will be prohibited by the CPU. Finally, the partition of multiple enclaves (rather than two) reduces the probability of interference between individual applications, thus protecting sensitive code and data against software attacks. For physical attacks, the integrated SE avoids bus data to be snooped. In the current design, the data stored in off-chip DDR is not protected against physical attacks since the DDR is not encrypted in the current implementation. XINE is also resistant to cache side-channel attacks because the EPA can flush the registers and caches during context switch. For power analysis, EM emissions, and fault injection attacks, XINE currently has no countermeasures. 

\section{Case Study} \label{sec:case-study}

As a case study, we choose a QR-code scanner. The QR-code scanner has been widely used in electronic payment and requires high-level security. The following steps lists a simplified process of an electronic payment.

\begin{itemize}
\item The embedded camera captures the QR-code image provided by the customer.
\item The embedded core or logic parses the QR-code image into binary data.
\item The SE encrypts the binary data and calculates its hash.
\item The QR-code scanner sends the encrypted data as well as its hash to the cloud via a secure channel.
\item The cloud verifies the authenticity and integrity of the data. Only if the data is authenticated and not modified, then the payment process continues.
\item Upon the payment is completed in cloud, the cloud notifies the QR-code scanner.
\end{itemize}

In this example, we design three App Enclaves. AE-1 is used for reading the QR image. AE-2 is used for parsing the QR image into binary data. If the parsed QR code is legal, then AE-2 requests the CE for operating authenticated encryption. The encrypted data, as well as its authentication tag, is then transferred to AE-3, which sends the encrypted data and its authentication tag to the cloud. AE-3 is also responsible for receiving response from the cloud. We note that AE-1 and AE-3 are used for communicating with external commands/data, and therefore both of them cannot affect the sensitive data and operation within AE-2.

\section{Summary} \label{sec:summary}

In this work, we propose XINE, a lightweight TEE framework for RISC-V based MCUs. The design follows the principles of leveraging security primitives, scalability, minimality, and extendibility. We employ the hardware primitive PMP to enforce isolation of applications. The isolation not only protects sensitive applications, but also jails malicious applications. We design special enclaves to control the SE exclusively and to store the runtime, respectively. In addition, we design a DMA for supporting efficient inter-enclave data transfer. As future work, we aim to extend XINE to more general scenarios, such as 1) a RISC-V core with M-, S-, and U-mode, 2) a multi-core microprocessor, and 3) a more complex SoC with bus masters other than the RISC-V core, and even with coprocessors.

\ifCLASSOPTIONcaptionsoff
  \newpage
\fi


%
\bibliographystyle{IEEEtran}   
\bibliography{IEEEabrv,xine}  

\end{document}